\documentclass[prb,floatfix,twocolumn,showpacs,amsfonts,amsmath,amssymb]{revtex4}
\usepackage{graphicx} 
\usepackage{dcolumn} 
\usepackage{bm} 
\usepackage{color}

\newcommand{\pii}{ \mbox{\boldmath{$\pi$}}}

\begin{document}
		
\title{G-factors of hole bound states in spherically symmetric
potentials  in cubic semiconductors}

\date{\today}

\author{D.~S.~Miserev, O.~P.~Sushkov} 
\affiliation{School of Physics, University of New South Wales, Sydney, Australia}

\begin{abstract}
Holes in cubic semiconductors have effective spin 3/2 and
very strong spin orbit interaction. Due to these factors properties
of hole bound states are highly unusual. We consider a single hole
bound by a spherically symmetric potential, this can be an acceptor or
a spherically symmetric quantum dot. 
Linear response to an external magnetic field is characterized by the
bound state Lande g-factor.
We calculate analytically g-factors of all bound states. 
\end{abstract}

\pacs{71.70.Ej, 
73.22.Dj, 
73.21.La, 
71.18.+y
}
\maketitle

\section{Introduction}
	Holes in cubic semiconductors originate from atomic p-orbitals.
Therefore, the total angular momentum of a hole can take values $j=3/2$
and $j=1/2$. So the spin orbit split valence band consists of 
$j=3/2$ and $j=1/2$ subbands.
In semiconductors with large fine structure the
$j=1/2$ state is very high in energy and therefore is irrelevant.
This is the case we consider here.
Following tradition in the field, below we call the angular momentum of the
hole ``the effective spin'' and denote it by letter $S$, $j\to S=3/2$.
Studies of holes in cubic semiconductors have a long history.
Pioneering theoretical work belongs to  Luttinger who has applied the
$k\cdot p$ method and the group theory analysis for hole quantum states 
near bottom of the band at the $\Gamma$ point~\cite{luttinger}.
Quadratic in momentum Luttinger Hamiltonian depends on
Luttinger parameters.  The Hamiltonian has a rotationally
invariant part which is independent of the lattice orientation and it also
contains a  small correction which is related to the cubic lattice and which is
not rotationally invariant.
In the present work we neglect the small 
rotationally
noninvariant correction and consider the rotationally invariant Hamiltonian.

The hole binding in an acceptor level or in a quantum dot is
a problem of a tremendous experimental interest.
Theoretically shallow acceptor levels have been studied thoroughly,
from the spherical approximation~\cite{baldereschi_sph} 
to accounting the lattice corrections~\cite{baldereschi_cube}, spin orbit split 
valence band~\cite{baldereshi_Si_Ge}, and so-called central cell 
correction~\cite{baldereshi_Si_Ge,fiorentini,drachenko}. 
As the acceptor spectrum is quite sensitive to variations of Luttinger 
parameters, the experimental data have been used to 
determine the parameters~\cite{Said}. The hole spectrum of the spherically symmetric quantum dot with infinite walls has also been calculated~\cite{xia}. 
	
Zeeman splitting of acceptor levels in an external magnetic field
has been also studied theoretically. The most detailed numerical calculation
includes the rotationally invariant approximation, 
cubic lattice correction, and also higher order magnetic field 
corrections~\cite{Schmitt}. Results of this calculation have been used
in analysis of numerous experimental data, see also~\cite{malyshev}.
Experimental data on the shallow acceptor spectrum in magnetic field have 
been obtained predominantly from  infrared absorption spectroscopy 
for different acceptors in 
GaAs~\cite{atzmuller,linnarsson,lewis_gaas}, 
Ge~\cite{fisher,baker,vickers,prakabar,wang}, Si~\cite{ramdas,kopf,heijden} and 
other semiconductors. 

We already pointed out above that in the present work we 
disregard the spin-orbit split $j=1/2$-band and limit our analysis
by the rotationally invariant approximation in the Luttinger Hamiltonian.
This implies that our analysis is applicable for GaAs, InAs, InSb, and 
on the other hand the approximation is not justified in Si. 
We calculate analytically g-factors of hole bound states for acceptors
and spherically symmetric single hole quantum dots.
At corresponding values of parameters our results are consistent
with that of previous numerical computations for acceptors~\cite{Schmitt}.

Structure of the paper is the following. In Section II we discuss the 
Hamiltonian and approximations. In Section III we calculate energy
levels and wave functions of a hole bound by a spherically symmetric
potential. This section is very similar to the paper~\cite{baldereschi_sph}.
Nevertheless we present this section because of two reasons, 
(i) it introduces the method and the notations which we use in following
sections.
(ii) We calculate previously unknown energy levels in a spherical parabolic
quantum dot.
Section IV presents our main result, analytical calculation of Zeeman splitting.
In Section V we specially consider the ``ultrarelativistic'' limit,
the limit when mass of the heavy hole is diverging.
Section VI presents tables of g-factors, and 
our conclusions are summarized in Section VII. 
Algebra of tensor operators is described in  Appendix A,
and the derivation of spin-orbit radial equations is given in Appendix B.

\section{Hamiltonian}
Luttinger Hamiltonian for $S=3/2$ free hole in a zinc-blende  semiconductor 
reads~\cite{luttinger}
\begin{eqnarray}
&&H_L = \left(\gamma_1 + \frac{5}{2} \gamma_2 \right) \frac{{\bf p}^2}{2 m} - \frac{\gamma_2}{m_0} \left(p_x^2 S_x^2 + p_y^2 S_y^2 + p_z^2 S_z^2  \right)
\nonumber \\
&&- \frac{\gamma_3}{m_0} \left(p_x p_y \{ S_x, S_y \} + p_y p_z \{ S_y, S_z \} + p_z p_x \{ S_z, S_x \} \right)\ , \nonumber
 \label{Lut}
\end{eqnarray}
x, y, z are the crystal axes of the cubic lattice,
${\bf p}$ is the quasi-momentum, $m$ is the electron mass,
$\gamma_1$, $\gamma_2$ and $\gamma_3$ 
are Luttinger parameters, and $\{... \}$ denotes the anticommutator. 
This Hamiltonian can be rewritten as
\begin{eqnarray}
H_L = \left(\gamma_1 + \frac{5}{2} \overline{\gamma}_2 \right)
\frac{{\bf p}^2}{2 m} 
- \frac{\overline{\gamma}_2}{m} \left({\bf p} \cdot {\bf S} \right)^2
+p_ip_jS_mS_nT^{(4)}_{ijmn}\ ,\nonumber
 \label{Lut1}
\end{eqnarray}
where
\begin{eqnarray}
\overline{\gamma}_2 = \frac{2 \gamma_2 + 3 \gamma_3}{5} \ .\nonumber
\end{eqnarray}
The irreducible 4th rank tensor $T^{(4)}_{ijmn}$ depends on the orientation 
of the cubic lattice, the tensor is proportional to $\gamma_2-\gamma_3$.
Since in the large
spin-orbit splitting materials $\gamma_2 \approx \gamma_3$ the rotationally
noninvariant part of the Hamiltonian can be neglected.
Hence the Luttinger Hamiltonian can be approximated by  the following
rotationally invariant (independent of the lattice orientation) Hamiltonian
\begin{eqnarray}
\label{LutSP}
H_L\to H_0 &=& \left(\gamma_1 + \frac{5}{2} {\overline \gamma}_2 \right) 
\frac{{\bf p}^2}{2 m} - \frac{{\overline \gamma}_2}{m} \left({\bf p} \cdot {\bf S} \right)^2\ .
\end{eqnarray}
Since different components of momentum commute between
themselves, $H_0$ can be rewritten as
\begin{eqnarray}
\label{LutSP1}
H_0 &=& \gamma_1 \frac{{\bf p}^2}{2 m} - \frac{{\overline \gamma}_2}{4 m} Q_{i j} \tau_{i j}
\\
\tau_{i j} &=& \left\{ S_i, S_j\right\} - \frac{2}{3} \delta_{i j} \cdot S (S+1)
\nonumber\\
Q_{i j}&=& \left\{ p_i, p_j\right\} - \frac{2}{3} \delta_{i j} \cdot p^2 
 \ .\nonumber
\end{eqnarray}
The $Q_{i j} \tau_{i j}$ term in (\ref{LutSP}) represents the effective spin
orbit interaction. Therefore, it is convenient to introduce the parameter
\begin{eqnarray}
\nu = \frac{2 \overline{\gamma}_2}{\gamma_1}\  , \ \ \ \nu < 1 \ ,\nonumber
\end{eqnarray}
which characterizes the relative strength of the spin orbit interaction.
Eigenstates of the Hamiltonian (\ref{LutSP}) have definite helicity,
the projection of spin on the direction of momentum, $S_p=\pm 1/2$ 
(light holes) and $S_p=\pm 3/2$ (heavy holes).
The corresponding masses are
\begin{eqnarray}
\label{mlh}
m_l=\frac{m}{\gamma_1(1+\nu)}\ , \ \ \
m_h=\frac{m}{\gamma_1(1-\nu)} \ .
\end{eqnarray}
In the limit $\nu \to 1$ mass of the heavy hole is diverging, $m_h \to \infty$,
therefore we call this the ``ultrarelativistic'' limit, the limit
corresponds to maximum possible spin orbit interaction.

To account for a magnetic field ${\bm B}$ imposed on the system
one has (i) to account for the vector potential via the
long derivative ${\bf p} \to \pii = {\bf p} - e {\bf A}$,
and (ii) to account for the Zeeman contribution $\propto {\bf S}\cdot{\bf B}$.
Here $e$ is the elementary electric charge.
Since different components of $\pii$ do not commute there is a well known 
ambiguity  between the gauge and the Zeeman terms.
In particular, 
$(\pii \cdot {\bf S})^2 = \frac{1}{4} \{\pi_i, \pi_j\} \cdot \{S_i, S_j\} 
- \frac{e}{2} {\bf B} \cdot {\bf S}$. 
Therefore the Zeeman term added to Eq.(\ref{LutSP}) 
is different from that added to (\ref{LutSP1}).
Here we follow the standard convention~\cite{Winkler} to use $H_0$ from
(\ref{LutSP1}) with
${\bf p} \to \pii = {\bf p} - e {\bf A}$.
In this case the spin Zeeman interaction is
\begin{equation}
\label{hz}
H_Z = - 2 \kappa \mu_B {\bf B} \cdot {\bf S},
\end{equation}
where $2 \kappa \mu_B {\bf S}$ is the spin magnetic moment,
$\mu_B$ is Bohr magneton, and value of $\kappa$ depends on the 
material~\cite{Winkler}. There is also cubic in spin Zeeman-like interaction
\begin{equation}
\label{hz1}
H_Z^{\prime} \propto S_iS_jS_mB_nT^{(4)}_{ijmn}\ ,
\end{equation}
however it is small~\cite{Winkler} and we disregard it.

We restrict our analysis by linear in magnetic field approximation.
Therefore, gauge terms, ${\bf p} \to \pii = {\bf p} - e {\bf A}$,
 also result in a magnetic moment.
As usually, see Ref.~\cite{Landau}, we use the gauge 
${\bf A} = [{\bf B} \times {\bf r}] /2$ to derive  the
orbital magnetic moment. 
All in all the operator of the total magnetic moment is
$\mu_B{\bf \mathfrak{M}}$, where
\begin{equation}
\mathfrak{M}_k = 2 \kappa S_k + \gamma_1 L_k - \gamma_1 \frac{\nu}{4} T_{i j k} \tau_{i j}.
\label{mag}
\end{equation}
Here the first term corresponds to (\ref{hz}), the second term
is the usual orbital magnetic moment which originates from the
$\gamma_1$-term in Eq.(\ref{LutSP1}), and the last term is a
spin dependent magnetic orbital moment which originates from the
$\gamma_2$-term in Eq.(\ref{LutSP1}). The 3d-rank tensor in the last term is
\begin{equation}
\label{t3}
T_{i j k} = \frac{1}{2} \left( \varepsilon_{i k n} \{ x_n, p_j\}  
+ \varepsilon_{j k n} \{ x_n, p_i\}\right) \ ,
\end{equation}
and $\tau_{ij}$ is defined in (\ref{LutSP1}).

Thus, the total Hamiltonian which we consider in this work is
\begin{equation}
\label{hu}
H=H_0+U(r)-\mu_B({\bf \mathfrak{M}}\cdot{\bf B})
\end{equation}
where $U(r)$ is spherically symmetric potential attracting the hole.

\section{Energy levels and wave functions of a hole in a  spherically symmetric
potential}
We already pointed out that this section is very similar to 
papers~\cite{baldereschi_sph,Schmitt}.
We present this section to introduce the method and the notations for
further calculations of g-factors.
Besides that we calculate previously unknown energy levels in a 
spherical quantum dot.
The Hamiltonian~(\ref{hu}) is invariant under rotations around the center
of symmetry of the potential. Therefore, while because of the 
spin orbit interaction the orbital angular momentum 
${\bf L} = [{\bf r} \times  {\bf p}]$ is  not conserved,
 the total angular momentum 
${\bf J} = {\bf L} + {\bf S}$ is conserved.
The eigenstates of the Hamiltonian (\ref{hu}) are also the 
eigenstates  of ${\bf J}^2=J(J+1)$ and $J_z=M$. 
The Hamiltonian contains only the orbital
scalar ${\bf p}^2$ and the orbital quadrupole $Q_{ij}$.
Therefore an eigenstate can be a combination of at most two values
of orbital angular momentum with $\Delta L=2$, we denote by $L$ the minimum 
of these two values
\begin{equation}
\Psi(L,\! S,\! J,\! M) = f(r) \cdot | L,\! S,\! J,\! M \rangle + g(r) \cdot |L+2,\! S,\! J,\! M \rangle,
\label{wavef}
\end{equation}
Radial wave functions are  normalized
$
\int\limits_{0}^{\infty} dr \, r^2 \left( |f|^2 + |g|^2 \right) = 1.
$
If $J \ge S = 3/2$, only two values of $L$ are possible: 
$L = J - 3/2$, $L = J - 1/2$. 
The special case of  $J = 1/2$ realizes only at $L = 1$ or $L = 2$. 
In this case $g(r) \equiv 0$ and the wave function is
\begin{equation}
\label{wavef1}
\Psi(L, S, J=1/2, M) = f(r) \cdot | L, S, J=1/2 , M \rangle,
\end{equation}
Interestingly, while the states with $J \ge S = 3/2$ are constructed of both
heavy and light holes, the states with $J=1/2$ are constructed of light
holes only. Masses of light and heavy holes behave differently in the
important limit $\nu \to 1$, see Eq.(\ref{mlh}).
We will see that this results in a qualitatively different behavior
of $J \ge 3/2$ and $J=1/2$ energy levels and g-factors in the
limit $\nu \to 1$.

To derive radial equations for functions $f$ and $g$ one has to use
Wigner-Eckart theorem and algebra of spherical tensor operators.
In Appendix A we remind the theorem and give definition of relevant
spherical tensors.
Equations for $f$ and $g$ are derived in Appendix B, here we just present the
equations. For $J=1/2$ there is only one equation (we set $\hbar=1$)
\begin{eqnarray}
\left[ - \frac{1}{2 m_l} \left( \partial_{rr} + \frac{2}{r} \partial_r - \frac{L(L+1)}{r^2}\right) + U(r) - E \right] f  =  0\nonumber\\
\label{jl}
\end{eqnarray}
We remind that here $L$  can only take  values $L = 1,2$.
\begin{widetext}
For $J\ge 3/2$ and  $L = J - 3/2$  the pair of coupled equations reads
\begin{eqnarray}
\begin{array}{ll}
\vspace{0.3 cm}
\displaystyle \left[ - \frac{\gamma_1 }{2 m} \left( 1 + \frac{\nu}{2} \frac{J - 3/2}{J} \right) \right.  &  \left. \left(\displaystyle \partial_{rr} + \frac{2}{r} \partial_r - \displaystyle \frac{(J - 1/2)(J - 3/2)}{r^2} \right) + U(r) - E \right] f + \\
\vspace{0.3 cm}
& + \displaystyle  \frac{\nu \gamma_1}{4 m} \frac{\sqrt{3(J - 1/2)(J + 3/2)}}{J} \left[ \partial_{rr} + \frac{2(J+1)}{r} \partial_r + \frac{(J - 1/2)(J + 3/2)}{r^2} \right] g =  0
\end{array} 
\nonumber
\label{radsyst11}
\\
\begin{array}{ll}
\vspace{0.3 cm}
\displaystyle \left[ - \frac{\gamma_1 }{2 m} \left( 1 - \frac{\nu}{2} \frac{J - 3/2}{J} \right) \right.  &  \left. \left(\displaystyle \partial_{rr} + \frac{2}{r} \partial_r - \displaystyle \frac{(J + 1/2)(J + 3/2)}{r^2} \right) + U(r) - E \right] g + \\
& + \displaystyle  \frac{\nu \gamma_1}{4 m} \frac{\sqrt{3(J - 1/2)(J + 3/2)}}{J} \left[ \partial_{rr} - \frac{2(J-1)}{r} \partial_r + \frac{(J + 1/2)(J - 3/2)}{r^2} \right] f =  0
\end{array} 
\end{eqnarray}
Finally, for $J \ge 3/2$ and  $L = J - 1/2$ the pair of coupled equations is
\begin{eqnarray}
\begin{array}{ll}
\vspace{0.3 cm}
\displaystyle \left[ - \frac{\gamma_1 }{2 m} \left( 1 - \frac{\nu}{2} \frac{J + 5/2}{J + 1} \right) \right.  &  \left. \left( \displaystyle \partial_{rr} + \frac{2}{r} \partial_r - \displaystyle \frac{(J - 1/2)(J + 1/2)}{r^2} \right) + U(r) - E \right] f + \\
\vspace{0.3 cm}
& + \displaystyle  \frac{\nu \gamma_1}{4 m} \frac{\sqrt{3(J - 1/2)(J + 3/2)}}{J + 1} \left[ \partial_{rr} + \frac{2(J+2)}{r} \partial_r + \frac{(J + 1/2)(J + 5/2)}{r^2} \right] g =  0
\end{array} 
\nonumber
\\
\begin{array}{ll}
\vspace{0.3 cm}
\displaystyle \left[ - \frac{\gamma_1 }{2 m} \left( 1 + \frac{\nu}{2} \frac{J + 5/2}{J+1} \right) \right.  &  \left. \left( \displaystyle \partial_{rr} + \frac{2}{r} \partial_r - \displaystyle \frac{(J + 3/2)(J + 5/2)}{r^2} \right) + U(r) - E \right] g + \\
& + \displaystyle  \frac{\nu \gamma_1}{4 m} \frac{\sqrt{3(J - 1/2)(J + 3/2)}}{J+1} \left[ \partial_{rr} - \frac{2 J}{r} \partial_r + \frac{(J - 1/2)(J + 3/2)}{r^2} \right] f =  0
\end{array} 
\label{radsyst12}
\end{eqnarray}
\end{widetext}
Unlike Eq.(\ref{jl}) which contains only mass of the light hole,
Eqs.(\ref{radsyst11}) and (\ref{radsyst12}) cannot be expressed solely in
terms of the heavy hole mass. Both heavy and light holes contribute
in $J\ge 3/2$ states. However, it is obvious that in the limit $\nu \to 1$,
$m_h \to \infty$, the heavy hole dominates in the wave function
and hence the energy must scale as
\begin{equation}
E \sim \frac{1}{m_h a^2},
\label{asymptE}
\end{equation}
where $a$ is the characteristic size of the wave function which also depends on
$\nu$. 
Following Ref.~\cite{baldereschi_sph} we will use the standard spectroscopic
notations for the quantum states, for example $1S_{3/2}$ indicates that $J=3/2$,
the principal quantum number is $n=1$, and $L=0$ (of course there is
also $L=2$ admixture).

\begin{figure}
	\begin{center}
		\includegraphics[angle = 0, width=0.4\textwidth]{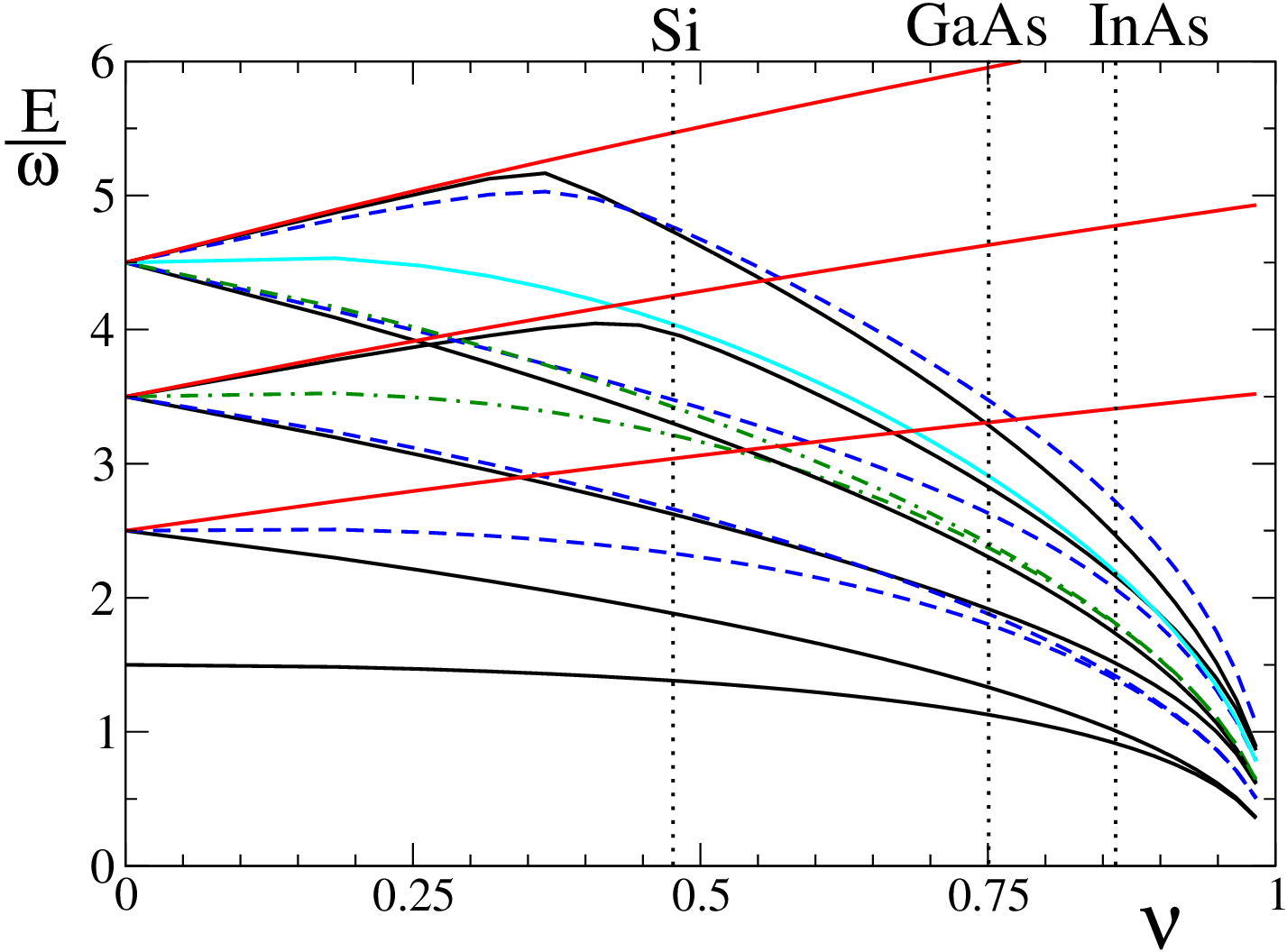}
		\caption{Energy levels versus $\nu$ in a parabolic quantum dot.
The levels with $J=1/2$ are shown by solid red lines, 
$2P_{1/2}$, $3D_{1/2}$, $4P_{1/2}$ from bottom to top.
The levels with $J=3/2$ are shown by solid black lines, 
$1S_{3/2}$, $2P_{3/2}$,  $3S_{3/2}$, $4P_{3/2}$, $3D_{3/2}$, $4F_{3/2}$
from bottom to top.
The levels with $J=5/2$ are shown by dashed blue lines,
$2P_{5/2}$, $3D_{5/2}$, $4P_{5/2}$,  $4F_{5/2}$
from bottom to top.
The levels with $J=7/2$ are shown by dashed-dotted green lines,
$3D_{7/2}$ and  $4F_{7/2}$ from bottom to top.
The level with $J=9/2$, $4F_{9/2}$ is shown by the solid cyan line.
Vertical dotted lines indicate values of $\nu$ that correspond to
particular compounds.
}
		\label{Spec_pic}
	\end{center}
\end{figure}

Numerical integration of Eqs.(\ref{jl}),(\ref{radsyst11}),(\ref{radsyst12})
is straightforward. Here we consider two  special cases, (i) parabolic
potential, and (ii) Coulomb potential.
The parabolic potential we take in the form
$$
U(r) = \frac{(m/\gamma_1) \omega^2 r^2}{2}.
$$
The ratio $E/\omega$ depends only on $\nu$.
Plots of $E/\omega$ versus $\nu$ for several lowest states are
presented in Fig.\ref{Spec_pic}.
The ground state is $1S_{3/2}$. While energies of states with $J=1/2$
grow with $\nu$ approximately linearly, the energies of states with
$J\ge 3/2$ scale as  $E\propto \sqrt{1-\nu}$ in agreement with
Eq.(\ref{asymptE}).
\begin{figure}
	\begin{center}
		\includegraphics[width=0.22\textwidth]{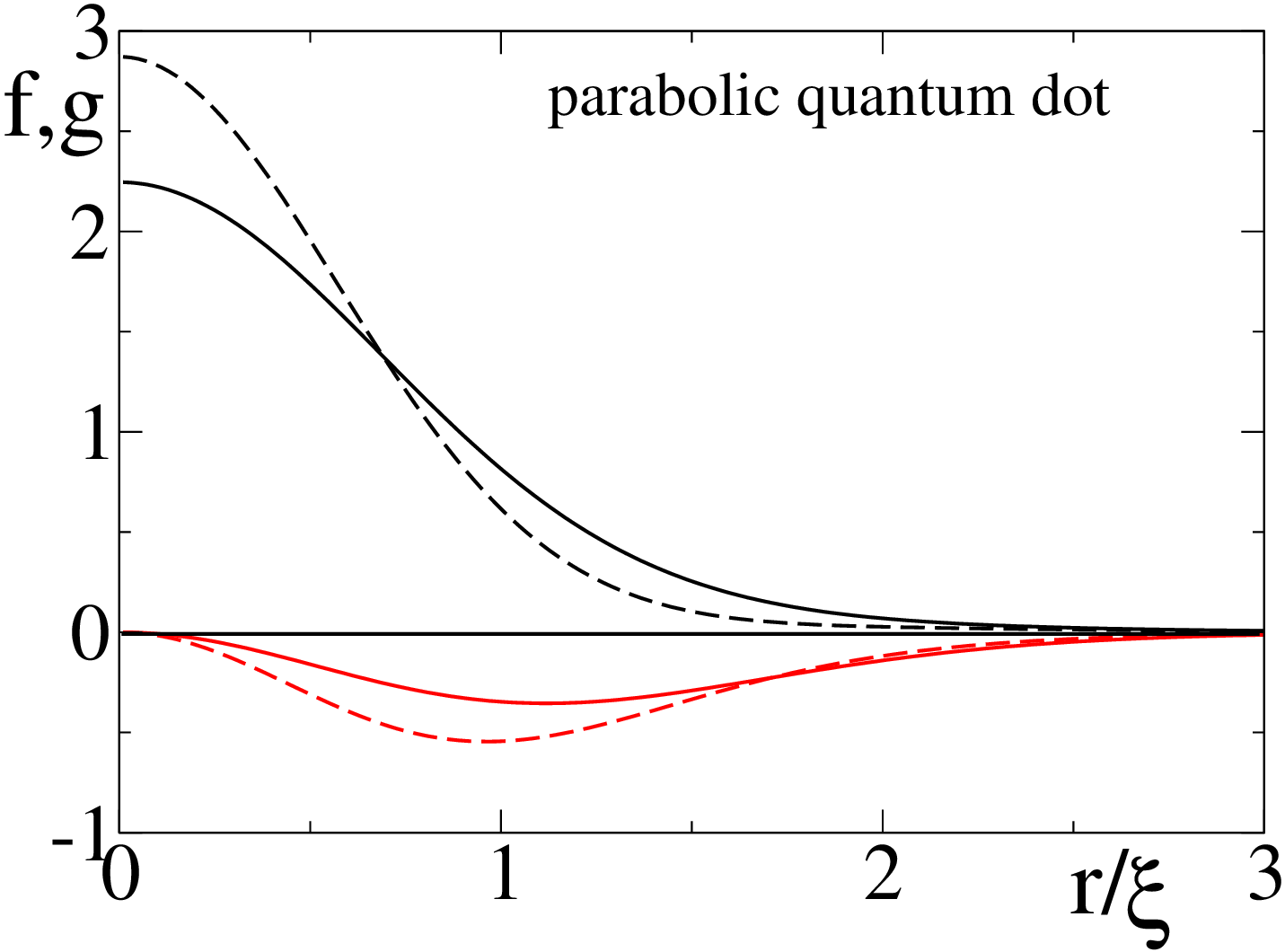}
\hspace{10pt}
		\includegraphics[width=0.22\textwidth]{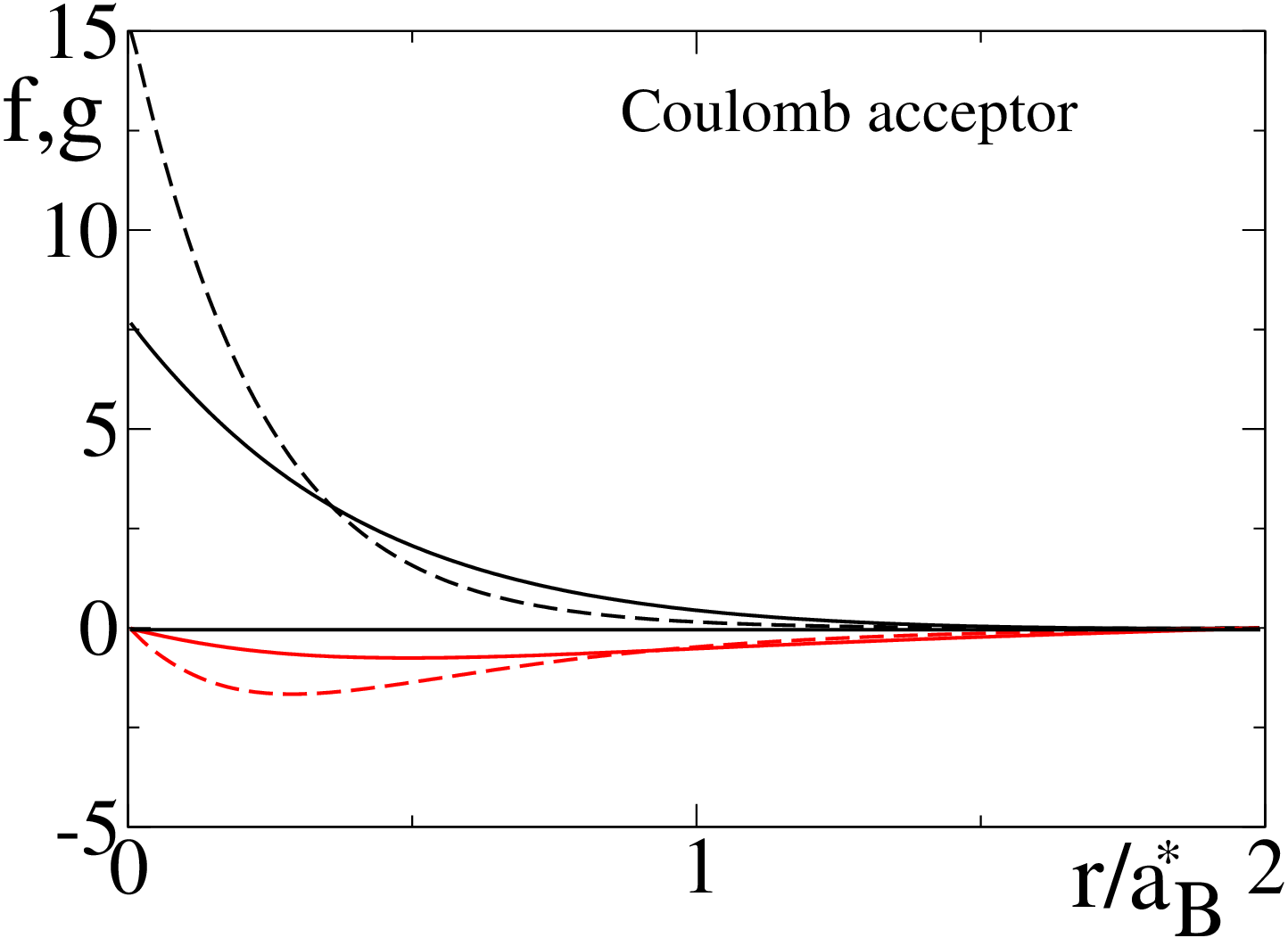}
		\caption{Radial wave functions 
of the ground state versus radius.
Black lines show $f(r)$ and red lines show $g(r)$. 
Solid lines correspond to $GaAs$ and dashed lines correspond to
$InAs$.
The left panel correspond to the parabolic quantum well. The scale is
$\xi=1/\sqrt{\omega m/\gamma_1}$.
The right panel corresponds to the Coulomb acceptor.
The scale is the effective Bohr radius,
$a_B^*=[(m/\gamma_1)(e^2/\epsilon)]^{-1}$.
}
		\label{Wfground}
	\end{center}
\end{figure}
For illustration we use Si, GaAs, and InAs. Values of Luttinger parameters
from Ref.\cite{Winkler} are presented in  Table \ref{t1L}.
\begin{table}[h]
	\begin{center}
		\begin{tabular}{|c|c|c|c|c|c|c|}
			\hline
  & $\gamma_1$ & $\gamma_2$ & $\gamma_3$ & ${\overline \gamma}_2$ &
$\nu$ &$\kappa$\\
			\hline

Si    &  4.285     &   0.339  &  1.446     & 1.00 & 0.47 &-0.42\\
GaAs  &  6.85      &  2.1     &   2.9      & 2.58 & 0.75 & 1.2\\
InAs  & 20.4       & 8.3      &  9.1       & 8.78 & 0.86 &7.6\\
			\hline
			\end{tabular}
	\end{center}
	\caption{Luttinger parameters for Si, GaAs, and InAs.}
	\label{t1L}
\end{table}
Our approach is valid when  $|\gamma_2-\gamma_3|$ is small compared to 
$\gamma_2$.
Obviously, the approach is not valid in Si, on the other hand it 
is quite reasonable in GaAs and even better in InAs.
 Values of $\nu$ corresponding to Si, GaAs, InAs are shown
in Fig.\ref{Spec_pic} by vertical dotted lines.
In the left panel of Fig.\ref{Wfground}
we plot radial wave functions of the quantum dot ground state for GaAs and InAs.
The s-wave is dominating, the weight of the d-wave,
\begin{equation}
\label{i1}
I_1 = \int\limits_{0}^{\infty} dr \, r^2 g^2
\end{equation}
is 21\% in GaAs and 32\% in InAs. 

\begin{figure}[ht]
\includegraphics[width=0.4\textwidth,clip]{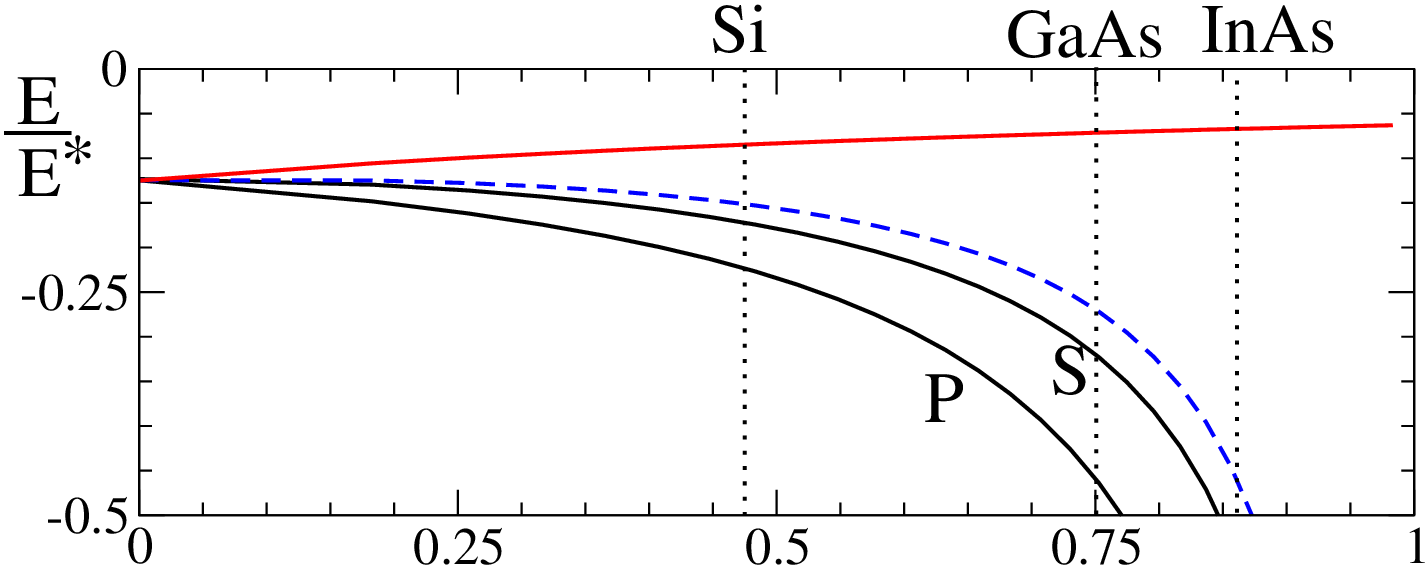}
\includegraphics[width=0.4\textwidth,clip]{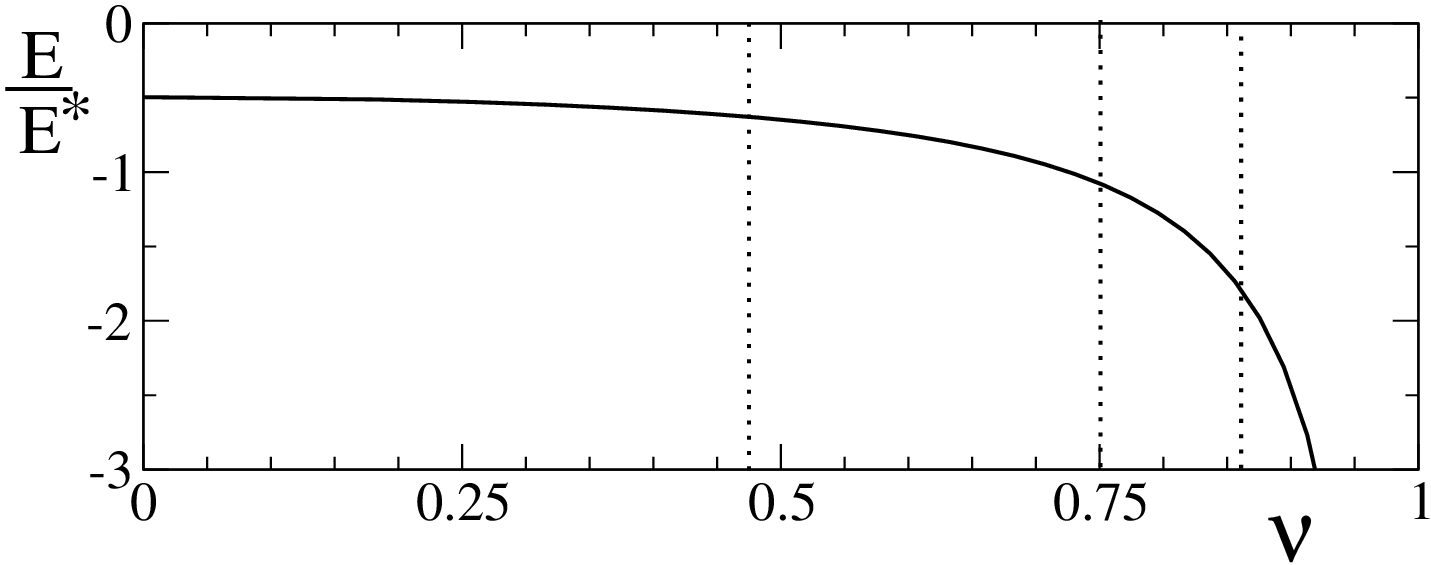}
\caption{Coulomb acceptor energy levels versus $\nu$.
The level with $J=1/2$, $2P_{1/2}$, is shown by the solid red line, 
top panel.
The levels with $J=3/2$ are shown by solid black lines, 
$1S_{3/2}$, in the bottom panel and $2P_{3/2}$,  $2S_{3/2}$,
in the top panel.
The level with $J=5/2$, $2P_{5/2}$, is shown by the dashed blue line,
top panel.
Note that the vertical axis scales are different in the top and in the
bottom panels.
Vertical dotted lines indicate values of $\nu$ that correspond to
particular compounds.
}
\label{SpeCoul}
\end{figure}

Another example is attracting Coulomb center corresponding to 
acceptor impurity.
$$
U(r) =-\frac{e^2/\epsilon}{r} \ ,
$$
where $\epsilon$ is the dielectric constant.
While this case has been thoroughly considered in the literature,
we briefly present the results since we use them in the next Section.
In this case the ``atomic energy unit'' is
$$
E^{*}= (m/\gamma_1) (e^2/\epsilon)^2 \ .
$$
The ratio $E/E^*$ is plotted in Fig.\ref{SpeCoul} versus $\nu$
for several lowest states.
In the right panel of Fig.\ref{Wfground} we plot the acceptor ground state
radial wave functions for GaAs and InAs.
Again, the s-wave is dominating, the weight of the d-wave, 
$I_1$,  is 26\% in GaAs and 35\% in InAs. 
Our results for the Coulomb acceptor are in excellent agreement with 
Ref.~\cite{Schmitt}.

\section{Lande g-factor}
The operator of magnetic moment $\mathfrak{M}$ is defined by Eq.(\ref{mag}).
Hence Lande g-factor of a quantum state with given $J$ is defined by
$$
\langle J M| \mathfrak{M}_z | J M \rangle  =g \cdot M \ .
$$
The operator $\mathfrak{M}$ contains  a simple spin, $\propto {\bf S}$,
 and a simple orbital, $\propto {\bf L}$, contribution,
as well as a more complex part dependent on the third rank tensor $T_{ijk}$,
where the tensor is defined by (\ref{t3}).
In order to apply the Wigner-Eckart formalism, we have to represent the third 
rank tensor in terms of irreducible tensors. It is easy to check that
\begin{eqnarray}
T_{i j k} = \frac{1}{3} \left( \varepsilon_{n i k} T^{(2)}_{j n}  + \varepsilon_{n j k} T^{(2)}_{i n}\right) - \nonumber\\ -\frac{1}{2} \left(\delta_{i k} L_j + \delta_{j k} L_i - 2 \delta_{i j } L_k\right),
\end{eqnarray}
where
\begin{equation}
T_{i j}^{(2)} = \frac{3}{4} \left( \{ x_i, p_j\}  + \{ x_j, p_i\}\right) - \frac{1}{2} \delta_{i j} \{ x_n, p_n\}
\end{equation}
is the irreducible tensor of the second rank.
For the magnetic moment (\ref{mag}) we need only the convolution $T_{i j k} \tau_{i j}$.
Therefore, we replace $T_{i j k}$ by the following simpler tensor
which gives the same convolution.
\begin{equation}
T_{i j k} \to \tilde{T}_{i j k} = \frac{2}{3} \varepsilon_{n j k} T^{(2)}_{i n} - \delta_{j k} L_i.
\end{equation}
Hence we can split  the magnetic moment operator (\ref{mag})
into four parts $\mathfrak{M}_z =  \mathfrak{M}_S + \mathfrak{M}_1 +\mathfrak{M}_2 +\mathfrak{M}_3$ 
which have different kinematic structures
\begin{eqnarray}
\label{mmmm}
&&\mathfrak{M}_S = 2 \kappa {\bf S}\nonumber\\
&&\mathfrak{M}_1 = \gamma_1 {\bf L}\nonumber\\
&&\mathfrak{M}_{2i} =\gamma_1 \frac{\nu}{4} \tau_{k i } L_k \nonumber\\
&&\mathfrak{M}_{3i} = - \gamma_1 \frac{\nu}{6}  \varepsilon_{n j i} 
T_{k n}^{(2)} \tau_{k j} \ .
\end{eqnarray}
The Lande factor splits correspondingly 
\begin{eqnarray}
\label{gggg}
&&g=g_s+g_1+g_2+g_3\\
&&g_s = 2 \kappa \frac{\langle  S_z \rangle}{M} \nonumber\\
&& g_1 = \gamma_1 \frac{\langle L_z \rangle}{M}\nonumber\\
&& g_2 =\gamma_1 \frac{\nu}{4} \frac{\langle \tau_{i z } L_i \rangle}{M}
 \nonumber \\
&&g_3 = - \gamma_1 \frac{\nu}{6}  
\frac{\langle \varepsilon_{n j z} T_{i n}^{(2)} \tau_{i j} \rangle}{M} \ ,
\nonumber
\end{eqnarray}
where we average over the state $|J M \rangle$
given by Eq.(\ref{wavef}) or Eq.(\ref{wavef1}).

\subsection{$\mathfrak{M}_S$  and $\mathfrak{M}_1$ contribution}
The first two contributions, $g_s$ and $g_1$, 
 in Eq.(\ref{gggg}) are usual textbook ones,
they can be easily calculated via ordinary vector model~\cite{Landau}:
\begin{eqnarray}
g_s \!=\! \kappa\! \left( 1 + \frac{S(S+1) -L(L+1) - 4 \left(L+\frac{3}{2}\right) I_1}{J(J+1)} \right)
\label{gs}
\end{eqnarray}
\begin{eqnarray}
g_1\! =\! \frac{\gamma_1}{2}\! \left(\!1+\!\frac{L(L+1) - S(S+1) + 4 \left(L+\frac{3}{2}\right) I_1}{J(J+1)} \right)
\label{g1}
\end{eqnarray}
Here $I_1$, Eq.(\ref{i1}), is the  weight of the component with higher 
orbital momentum $L+2$. Notice that $I_1=0$ for $J=1/2$.

\subsection{$\mathfrak{M}_2$ contribution}
In Eqs.(\ref{mmmm}),(\ref{gggg})
 the operator $\mathfrak{M}_2$ is defined as product of
Cartesian tensors, $\mathfrak{M}_{2i} \propto D_i = \tau_{i k} L_k$.
On the other hand the tensor product theorem (\ref{tenpro}) is formulated
in terms of spherical tensors where the product is defined in terms of
Clebsch-Gordon coefficients $C^{\alpha\beta}_{\gamma\delta,\xi\zeta}$
\begin{equation}
\label{D}
D^{(1)}_q = \sum \limits_{q'} C^{1 q}_{1 q'  2\, q-q'} L^{(1)}_{q'} \tau^{(2)}_{q-q'} \ .
\end{equation}
For $L_i$ and $\tau_{ij}$ we use the conventions (\ref{vector}), 
(\ref{tensharm2}) to relate spherical and Cartesian components.
Hence the relation between Cartesian and spherical components of $D$ is 
determined by (\ref{D}).
\begin{eqnarray}
D^{(1)}_0= C^{10}_{10,20}D_z=- \sqrt{\frac{2}{5}} D_z \ .\nonumber
\end{eqnarray}
Now we can use the general tensor product formula (\ref{tenpro}) in combination
with Wigner-Eckart theorem to find the matrix element of $D_z$
\begin{eqnarray}
&&\langle L', S, J, M| D_z | L, S, J, M \rangle = -M \sqrt{\frac{15}{2} \frac{2J+1}{J(J+1)}} \times \nonumber\\ 
&&\ \ \ \times \left\{\begin{array}{ccc}
1 & 2 & 1\\
L & S & J\\
L & S & J
\end{array}\right\} \langle L|| L^{(1)} || L \rangle \cdot \langle S || \tau^{(2)} || S \rangle \delta_{L L'},
\label{Delem}
\end{eqnarray}
where $\{...\}$ is the $9j$-symbol and the reduced matrix elements are
\begin{eqnarray}
\label{rm}
\langle L|| L^{(1)} || L \rangle &=& \sqrt{L(L+1)(2L+1)}\nonumber\\
\langle S || \tau^{(2)} || S \rangle &=& \frac{1}{3} \sqrt{(2S-1)2S(2S+1)(2S+2)(2S+3)}\nonumber\\
&=& 4 \sqrt{5} \ .
\end{eqnarray}
In Eq.(\ref{Delem}) we keep in mind that $f$ and $g$-components of the 
wave function differ by $\Delta L=2$, and the operator $D_i$ which
is linear in
 $L$, does not connect these components.
Hence, the $g_2$-Lande factor takes the form:
\begin{eqnarray}
g_2 &&
= \gamma_1 \frac{ \nu}{4 M} \langle \Psi | D_z | \Psi \rangle  \nonumber\\
&&= \gamma_1 \frac{ \nu}{4 M} \left( \vphantom{\frac{1}{2}} \langle L, S, J, M| D_z | L, S, J, M \rangle \cdot (1 - I_1) + \right. \nonumber\\
&& \ \ \ \ \left. \langle L+2, S, J, M| D_z | L+2, S, J, M \rangle \cdot I_1\right),
\end{eqnarray}
where the matrix elements are given by (\ref{Delem}). 
Substituting $9j$-symbols we arrive to the following answer
\begin{eqnarray}
\label{g2}
&&J=1/2:\nonumber\\
&&g_2= - \gamma_1 \frac{5\nu}{3}\ \ \  if  \ \ L=1\nonumber\\
&&g_2=  \gamma_1 \nu  \ \ \ \ \ \ \      if   \ \  L=2\nonumber\\
&&\\
&& J \ge 3/2:\nonumber\\
&&g_2 = \gamma_1 \frac{\nu}{2} \left( \frac{J- 3/2}{J} - \frac{2 J +1 }{J+1} I_1 \right)\  \ \ \ \ \ \ \ if \  L = J - 3/2\nonumber\\
&&g_2 = \gamma_1 \frac{\nu}{2} \left( \frac{J+ 5/2}{J+1} - \frac{2 J +1}{J} (1- I_1) \right) \ if \ L = J - 1/2 \nonumber
\end{eqnarray}

\subsection{$\mathfrak{M}_3$ contribution}
The calculation of the  $\mathfrak{M}_3$ contribution is
very similar to $\mathfrak{M}_2$.
The operator $\mathfrak{M}_3$, Eqs.(\ref{mmmm}),(\ref{gggg}),
is defined as vector product of two irreducible Cartesian tensors 
$\tau$ and $T^{(2)}$ of  second rank, 
$\mathfrak{M}_{3i} \propto R_i = \varepsilon_{nij}T^{(2)}_{kn}\tau_{kj}$.
On the other hand the tensor product theorem (\ref{tenpro}) is formulated
in terms of spherical tensors where the product is defined in terms of
Clebsch-Gordon coefficients
\begin{equation}
\label{R}
R^{(1)}_q = \left( \tau^{(2)} \times T^{(2)} \right)_{1 q} = \sum \limits_{q'} C^{1 q}_{2 q'  2\, q-q'} \tau^{(2)}_{q'} T^{(2)}_{q-q'}.
\end{equation}
For $T^{(2)}_{ij}$ and $\tau_{ij}$ we use the convention 
(\ref{tensharm2}) to relate spherical and Cartesian components.
Hence the relation between Cartesian and spherical components of $R_i$ is 
determined by (\ref{R}).
\begin{equation}
R_0^{(1)} = \frac{4 i}{3}C^{10}_{2,-1;2,1} R_z=\frac{4 i}{3 \sqrt{10}} R_z.
\end{equation} 
According to~(\ref{tenpro}), matrix elements of $R_0^{(1)}$ break into the product of reduced matrix elements for $T^{(2)}$ and $\tau$:
\begin{eqnarray}
\langle L', S, J, M| R^{(1)}_0 | L, S, J, M \rangle = M \sqrt{\frac{3( 2J+1)}{J(J+1)}} \times \nonumber \\ \times \left\{\begin{array}{ccc}
2 & 2 & 1\\
S & L' & J\\
S & L & J
\end{array}\right\} \langle L'|| T^{(2)} || L \rangle \cdot \langle S || \tau^{(2)} || S \rangle.
\end{eqnarray}
The reduced matrix element of $\tau^{(2)}$ is given in (\ref{rm}).
Non-zero reduced matrix elements for $T^{(2)}$ are
	\begin{eqnarray}
&&\langle L|| T^{(2)} || L \rangle = - \sqrt{\frac{L(L+1)(2L+1)}{(2L - 1)(2L+3)}}\nonumber\\
&&\hspace{53pt} \times\left( \vphantom{\frac{1}{2}} \{r, p_r\} - 2 i \right)\nonumber \\
\label{t22}
&&\langle L + 2|| T^{(2)} || L \rangle = \sqrt{\frac{3(L+1)(L+2)}{2(2L+3)}}\\ 
&&\hspace{70pt} \times\left( \vphantom{\frac{1}{2}} \{r, p_r\} + i (2L+1) \right)\nonumber\\
&&	\langle L - 2|| T^{(2)} || L \rangle = \sqrt{\frac{3L(L-1)}{2(2L-1)}}
\nonumber\\
&&\hspace{70pt} \times\left( \vphantom{\frac{1}{2}} \{r, p_r\} - i (2L+1) \right),\nonumber
	\end{eqnarray}
where $p_r = - i \partial_r$.
The diagonal matrix element $\langle L|| T^{(2)} || L  \rangle$ gives zero
after the r-integration, so there are no $ff$ and $gg$ terms  in 
$\mathfrak{M}_3$. 
This is a direct consequence of the time reversal invariance,
$T^{(2)}$ changes sign under time reversal and on the other hand it is
impossible to construct a second rank tensor which is consistent
with Wigner-Eckart and which changes sign under time reversal.
This implies that $g_3$ is zero for $J=1/2$ states.
However, the $fg$ contribution to $\mathfrak{M}_3$ which originate from
(\ref{t22}) is nonzero. A direct calculation with (\ref{t22}) gives the 
following answer
\begin{eqnarray}
\label{g3}
&&g_3 = i \gamma_1 \sqrt{\frac{5}{2}} \cdot \frac{\nu}{4 M} \cdot \langle \Psi| R^{(1)}_0 | \Psi \rangle  \\ 
&&\ \ \ = \gamma_1 \nu \frac{\sqrt{3 (J -1/2) (J+ 3/2)}}{2 J (J+1)} \!\!
\left[ 2 \left( L + \frac{3}{2} \right) I_2 + I_3 \right]\ ,\nonumber 
\end{eqnarray}
where
\begin{eqnarray}
\label{ig3}
&&I_2 =  \int\limits_{0}^{\infty} dr \, r^2 f(r) g(r)\nonumber \\
&&I_3 =  \int\limits_{0}^{\infty} dr \, r^3 (f g' - g f')\ .
\end{eqnarray}
We remind our convention,  $L$ is the minimum of two mixing orbital momenta,
$L$ corresponds to $f$ and $L+2$ corresponds to $g$.

Eqs. (\ref{gggg}), (\ref{gs}), (\ref{g1}), (\ref{g2}), and (\ref{g3})
solve the problem of g-factor of the hole bound state.
Answers for $J=1/2$ states are universal
\begin{eqnarray}
\label{g12}
&&P_{1/2}: \ \ \ g=\frac{10}{3}\kappa-\frac{\gamma_1}{3}(2+5\nu)\nonumber\\
&&D_{1/2}: \ \ \ g=-2\kappa+\gamma_1(2+\nu)\ .
\end{eqnarray}
Answers for $J\ge 3/2$ states are not universal, they depend
on radial wave functions. In Section VI we present the results for
parabolic quantum dot.
However, before calculating numerical values of the g-factor  we 
discuss the ``ultrarelativistic'' limit $\nu \to 1$.

\section{g-factor in ``ultrarelativistic'' $\nu \to 1$ limit}
The case of maximum possible spin-orbit coupling $\nu \to 1$ 
($\gamma_1 = 2 \gamma_2$) is important as a theoretical limit and also
 some real materials such as InSb and InAs are
pretty close to this limit.
As it has been mentioned above, energies of $J \ge 3/2$ states 
in the limit $\nu \to 1$ satisfy the approximate relation~(\ref{asymptE}).
The states fall down on the bottom of the well.
On the other hand, the states
are described by equations (\ref{radsyst11}) and (\ref{radsyst12}).
Because of collapse of the wave functions 
 at $\nu=1$ we can neglect the energy and the potential 
in these Eqs. We keep only $\partial_{rr}$, $\frac{1}{r}\partial_{r}$ 
and $\frac{1}{r^2}$ terms. Excluding the second derivatives by subtraction of two coupled radial equations,
 we find the following relation
 between functions $f(r)$ and $g(r)$:
\begin{eqnarray}
\label{fgmu1}
&&\alpha \left(f' - \frac{L}{r}f \right) = g' + \frac{L+3}{r} g\\
&&\alpha = \left\{
\begin{array}{ccc}
\sqrt{\frac{3(J-1/2)}{J+3/2}}, \quad L = J-3/2, \\
\sqrt{\frac{J-1/2}{3(J+3/2)}}, \quad L = J -1/2.
\end{array} \right. \nonumber
\end{eqnarray}
Multiplying Eq.(\ref{fgmu1}) by $gr^3$ and separately by $fr^3$ and then
integrating by parts keeping in mind the normalization condition,
$\int f^2r^2dr=1-I_1$, one finds the following universal expression for the
weight of the $g$-component in the ``ultrarelativistic'' limit
\begin{eqnarray}
\label{i1r}
I_1 = \left\{
\begin{array}{ll}
3(J-1/2)/(4J), & L =J -3/2,\\
(J-1/2)/(4(J+1)), & L=J-1/2.
\end{array}\right.
\end{eqnarray}
The $g_3$-contribution (\ref{g3}) depends on the radial integrals $I_2$ and
$I_3$ defined in Eq.(\ref{ig3}).
The relation (\ref{fgmu1}) is not sufficient to determine $I_2$ and
$I_3$ separately. However, remarkably, the combination
$2(L+3/2)I_2+I_3$ which enter Eq. (\ref{g3}) can be found from (\ref{fgmu1}).
Using the same derivation as (\ref{i1r}) we find
\begin{equation}
\label{i23r}
2(L+3/2)I_2+I_3 = -\frac{1}{2}\sqrt{3\left(J - \frac{1}{2}\right)\left(J + \frac{3}{2}\right)} \ .
\end{equation}
Substitution of the integrals (\ref{i1r}) and (\ref{i23r}) in
Eqs. (\ref{gggg}), (\ref{gs}), (\ref{g1}), (\ref{g2}), and (\ref{g3})
gives the following values of the g-factors for $J \ge 3/2$
\begin{eqnarray}
\label{5g}
&&g_s = \frac{9 \kappa}{2J(J+1)}\nonumber \\
&&g_1 = \gamma_1\left( 1 - \frac{9}{4J(J+1)}\right)\nonumber \\
&&g_2 = -\frac{\gamma_1}{4}\left( 1+ \frac{9}{4J(J+1)} \right)\nonumber \\
&&g_3 = -\frac{3 \gamma_1}{4} \left( 1 - \frac{3}{4J(J+1)}\right)\nonumber \\
&&g = \frac{9(2\kappa - \gamma_1)}{4J(J+1)} \ .
\end{eqnarray}
Interestingly, Eqs.(\ref{5g}) are the same for $L=J-3/2$ and $L=J-1/2$.
Thus in the ``ultrarelativistic'' limit, $\nu \to 1$,  $g$-factors 
of $J\ge 3/2$ states depend only  on $J$
and they are independent of a particular confinement (parabolic
quantum dot, Coulomb
acceptor,...) and independent of the radial quantum number.
This universal behaviour is due to the collapse of heavy hole
wave functions. Another point to note is the $\propto 1/[J(J+1)]$
dependence of the g-factor (\ref{5g}).
This is similar to the g-factor dependence in diatomic molecules~\cite{Landau}.
Physical reasons for this are also very similar, in the molecules the
dependence is due to the spin-axis interaction, for holes the
dependence is due to $\left({\bf p} \cdot {\bf S} \right)^2$ spin-momentum Luttinger interaction.

\section{Numerical values of the Lande factor}
We have already pointed out that, in principle, Eqs. (\ref{gggg}), (\ref{gs}), 
(\ref{g1}), (\ref{g2}), (\ref{g3})
solve the problem of g-factor of the hole bound state.
Still, to calculate numerical values of g-factors one needs to calculate
integrals $I_1$, $I_2$, and  $I_3$. The integrals immediately follow from wave
functions found in Section III.
We have checked that our results for g-factors of Coulomb acceptor states
perfectly agree with that found numerically and tabulated in 
Ref.~\cite{Schmitt}. Below we present our results for three lowest states
in parabolic quantum dot, $1S_{3/2}$, $2P_{3/2}$, $2P_{5/2}$, Tables
\ref{table_parab1}, \ref{table_parab2}, and \ref{table_parab3}
respectively.
We tabulate  
$g_s/\kappa$, $g_1/\gamma_1$, $g_2/\gamma_1$, $g_3/\gamma_1$
for different values of $\nu$. 
In the tables we also present separately the total orbital contribution $g_L$
\begin{eqnarray}
g_L = g_1+g_2+g_3, \quad g = g_s + g_L \ . \nonumber
\end{eqnarray}
All lines in each table except of the lowest line are calculated 
with Eqs. (\ref{gggg}), (\ref{gs}), (\ref{g1}), (\ref{g2}), (\ref{g3})
and with integrals $I_1$, $I_2$,  $I_3$ calculated with 
functions $f $ and $g$ found numerically in Section III.
The lowest lines are calculated directly with the ``ultrarelativistic''
Eqs.(\ref{5g}).

\begin{table}[h]
	\begin{center}
		\begin{tabular}{|c|c|c|c|c|c|}
			\hline
			$\nu$ & $g_s/\kappa$ & $g_1/\gamma_1$ & $g_2/\gamma_1$ & $g_3/\gamma_1$ & $g_L/\gamma_1$ \\ 
			\hline
			0.6& 1.83& 0.087& -0.052& -0.285& -0.25 \\
			0.65& 1.78& 0.109& -0.071& -0.337& -0.299 \\
			0.7& 1.73& 0.135& -0.095& -0.392& -0.352 \\
			0.75& 1.67& 0.166& -0.125& -0.449& -0.408 \\
			0.8& 1.59& 0.203& -0.162& -0.505& -0.465 \\
			0.85& 1.51& 0.246& -0.209& -0.556& -0.519 \\
			0.9& 1.408& 0.296& -0.266& -0.596& -0.566 \\
			0.95& 1.3& 0.349& -0.332& -0.615& -0.597 \\
			1.0& 1.2& 0.4& -0.4& -0.592& -0.6 \\
			\hline
			\end{tabular}
	\end{center}
\vspace{-10pt}
	\caption{g-factor contributions for $1S_{3/2}$ ground state in the 
parabolic quantum well at different values of $\nu$.} 
	\label{table_parab1}
\end{table}
\begin{table}[h]
	\begin{center}
		\begin{tabular}{|c|c|c|c|c|c|}
			\hline
		$\nu$ & $g_s/\kappa$ & $g_1/\gamma_1$ & $g_2/\gamma_1$ & $g_3/\gamma_1$ & $g_L/\gamma_1$ \\ 
			\hline
		0.6& 1.375& 0.312& -0.293& -0.251& -0.231 \\ 
		0.65& 1.362& 0.319& -0.313& -0.288& -0.282 \\  
		0.7& 1.347& 0.326& -0.332& -0.327& -0.333 \\ 
		0.75& 1.331& 0.334& -0.349& -0.369& -0.384 \\ 
		0.8& 1.314& 0.343& -0.365& -0.412& -0.434 \\ 
		0.85& 1.294& 0.353& -0.380& -0.457& -0.484 \\ 
		0.9& 1.271& 0.365& -0.392& -0.504& -0.532 \\ 
		0.95& 1.243& 0.379& -0.400& -0.553& -0.574 \\ 
		1.0& 1.2& 0.4& -0.4& -0.592& -0.6 \\
			\hline
		\end{tabular}
	\end{center}
\vspace{-10pt}
	\caption{g-factor contributions for the first excited $2P_{3/2}$ 
state in the
parabolic quantum well at different values of $\nu$.} 
	\label{table_parab2}
\end{table}
\begin{table}[h]
	\begin{center}
		\begin{tabular}{|c|c|c|c|c|c|}
			\hline
			$\nu$ & $g_s/\kappa$ & $g_1/\gamma_1$ & $g_2/\gamma_1$ & $g_3/\gamma_1$ & $g_L/\gamma_1$  \\ 
			\hline
			0.6& 0.897& 0.552& -0.016& -0.427& 0.108 \\ 
			0.65& 0.841& 0.579& -0.045& -0.483& 0.052 \\ 
			0.7& 0.784& 0.608& -0.078& -0.534& -0.004 \\ 
			0.75& 0.728& 0.636& -0.116& -0.578& -0.058 \\ 
			0.8& 0.674& 0.663& -0.155& -0.615& -0.108 \\
			0.85& 0.625& 0.687& -0.197& -0.644& -0.153 \\ 
			0.9& 0.582& 0.709& -0.237& -0.665& -0.193 \\ 
			0.95& 0.545& 0.728& -0.277& -0.679& -0.228 \\ 
			1.0& 0.514& 0.743& -0.314& -0.677& -0.257 \\ 
			\hline
		\end{tabular}
	\end{center}
\vspace{-10pt}
	\caption{g-factor contributions for the second excited $2P_{5/2}$ 
state in the parabolic 
quantum well at different values of $\nu$.} 
	\label{table_parab3}
\end{table}
In all the cases there is a very strong compensation between spin, $g_s$, and 
orbital, $g_{L}$, contributions to the g-factor.
For example, in GaAs ($\nu=0.75$), $g_{1S_{3/2}}=2.0-2.8=-0.8$,
$g_{2P_{3/2}}=1.6-2.6=-1.0$, $g_{2P_{5/2}}=0.87-0.40=+0.47$.
These compensations are somewhat similar to the compensation found
in a particular situation in Ref.~\cite{Uli}.
Here we see that the compensation is a generic effect related to the
``ultrarelativistic'' formula (\ref{5g}).

\section{conclusions}
We have developed analytical tensor theory of Lande g-factors 
in spherically symmetric bound states of holes
in cubic semiconductors. The Lande factors are given
by Eqs. (\ref{gggg}), (\ref{gs}), (\ref{g1}), (\ref{g2}), (\ref{g3})
with integrals $I_1$, $I_2$, $I_3$ which depend on the quantum state
wave function, see Eqs. (\ref{wavef}), (\ref{i1}), (\ref{ig3}).
We have shown that g-factors only weakly depend on the type
of confinement (Coulomb acceptor, parabolic quantum dot, ....).
Behaviour of g-factors in the ``ultrarelativistic'' limit is
universal and completely independent of the type of confinement.
This ``ultrarelativistic'' behaviour enforces  a strong compensation
between spin and orbital contributions to g-factors.
It is worth noting that the developed tensor technique provides
a solid basis for calculation of g-factor of asymmetric quantum dots,
this problem will be addresses separately.\\

\textbf{Acknowledgements}
We acknowledge Alex Hamilton, Tommy Li, Qingwen Wang,
Ulrich Zuelicke and Roland Winkler for important discussions and interest to the
work.\\

\appendix
\section{irreducible tensors, Wigner-Eckart and tensor product theorems}

In this Appendix we summarize the major equations from the
SO(3) tensor algebra~\cite{Landau} which we use in our calculations.
All tensors are defined in three-dimensional space.
A tensor $R_{i_1 i_2 \dots i_\ell}$ of the rank $\ell$ is irreducible if its 
components can be transformed over some irreducible matrix representation 
$\mathfrak{D}_\ell$ of the rotational group  SO(3). 
Consequently, the number of independent components of $R_{i_1 i_2 \dots i_\ell}$ 
coincides with the dimension of the representation 
$\dim \mathfrak{D}_\ell = 2 \ell +1$, whence we may choose the basis of 
these components $R_m^{(\ell)}$, $m = -\ell, \dots, \ell$ which have been 
transformed exactly like spherical harmonics $Y_{\ell m}$. This basis is 
usually called spherical basis. Spherical components of the vector ${\bf R}$
can be related to its Cartesian components in the following way
\begin{eqnarray}
R_0^{(1)} = R_z, \quad R_{\pm 1}^{(1)} = \mp \sqrt{\frac{1}{2}} (R_x \pm iR_y) \ .
\label{vector}
\end{eqnarray}
Similarly spherical harmonics of the second rank irreducible tensor $
R^{(2)}$ can be related to its Cartesian components as
\begin{eqnarray}
R_0^{(2)} = R_{z z}, \; R^{(2)}_{\pm 1} = \mp \sqrt{\frac{2}{3}} \left( R_{x z} \pm i R_{y z}  \right), \nonumber \\ R^{(2)}_{\pm 2} = \sqrt{\frac{1}{6}} \left( R_{x x} - R_{y y} \pm 2 i R_{x y} \right),
\label{tensharm2}
\end{eqnarray}

The Wigner-Eckart theorem reduces the matrix element of a spherical tensor
between states with given angular momentum $|J,J_z=M\rangle$ to the 
3j-symbol~\cite{Landau}
\begin{equation}
\langle J'\!,\! M'| R^{(\ell)}_\kappa | J,\! M \rangle \! = \!(-1)^{J'\! -\! M'} \!\! 
\left(\begin{array}{ccc}
\! \!J'\!\! &\! \! \ell \! &\! \! J \!\!\\
\! \!-M'\! &\! \! \kappa \! & \!M\!\!
\end{array}\right) \!\! \langle J'|| R^{(\ell)} || J \rangle,
\label{wigner}
\end{equation}
Importantly, the reduced matrix element $\langle J'|| R^{(\ell)} || J \rangle$
is independent of the projections $M$ and $M'$.

Consider tensor product of two irreducible tensors $T^{(k)}$ and $Q^{(m)}$ of 
ranks $k$ and $m$ respectively. The tensor product is also an irreducible
tensor of rank $|k-m| \le \zeta \le k+m$. It is defined as
\begin{eqnarray}
\hspace{-10pt}R^{(\zeta)}_q = (T^{(k)}\times Q^{(m)})_{\zeta q} = \sum_{q'} C^{\zeta q}_{k\, q', m\, q-q'} T^{(k)}_{q'} Q^{(m)}_{q-q'}
\end{eqnarray}
where $C$ is the Clebsch-Gordon coefficient. 
The scalar product is a special case of the tensor product
\begin{equation}
(T^{(k)} \cdot Q^{(k)}) = \sum\limits_{q = -k}^{k} (-1)^q Q^{(k)}_q T^{(k)}_{-q}.
\label{scalar}
\end{equation}

If two irreducible tensors $T$ and $Q$ in the tensor product
are independent, i.e. they act in different sub-spaces (say spin and orbit)
then the matrix element of the tensor product is
disentangled via the product of reduced matrix elements of $T$ and $Q$:
\begin{widetext}
	\begin{eqnarray}
&&\hspace{-20pt}\langle J_1' J_2' J' M'| (T^{(k)}\! \!\times Q^{(m)} )_{\zeta q} | J_1 J_2 J M \rangle \!\!  =  \! \!(-1)^{J'-M'}\! \Pi_{\zeta,J,J'} \!\! \left(\begin{array}{ccc}
	\!J'\! & \!\zeta \! &\! J \!\\
	\!-M'\! &\! q \!&\! M \!
	\end{array}\right) \!\!
	\left\{\begin{array}{ccc}
	\!\!k &\! m \!& \!\zeta \! \!\\
	\!\!J_1' & \! J_2'\! &\! J' \! \!\\
	\!\!J_1 & \!J_2 \!&\! J \!\!
	\end{array}\right\} \!
	\langle J_1'|| T^{(k)} || J_1 \rangle \! \langle J_2'|| Q^{(m)} || J_2 \rangle
	\label{tenpro} \\
&&\hspace{-20pt}\langle J_1' J_2' J' M'| (T^{(k)}\cdot Q^{(k)}) | J_1 J_2 J M \rangle  =  \delta_{J' J} \cdot \delta_{M' M} (-1)^{ J+ J_1 + J_2'} \left\{\begin{array}{ccc}
	J_1' & J_1 & k\\
	J_2 & J_2' & J
	\end{array}\right\} \langle J_1'|| T^{(k)} || J_1 \rangle \langle J_2'|| Q^{(k)} || J_2 \rangle \ ,
	\label{qtau}
	\end{eqnarray}
\end{widetext}
where $\Pi_{\zeta,J,J'} = \sqrt{(2 \zeta+1)(2 J+1)(2 J'+1)}$, 
and curly braces denote $9 j$ and $6 j$ symbols.

\section{spin-orbit radial equations}
To derive equations for radial functions of the bound hole
we have to calculate matrix elements from tensors $\tau^{(2)}$ and $Q^{(2)}$ 
defined in (\ref{LutSP1}). Since the general dependence of projections
is given by the Wigner-Eckart theorem, Eq. (\ref{wigner}), we need to know
only reduced matrix elements. Let us  calculate for example the diagonal
matrix element $\langle L|| Q^{(2)} || L \rangle $
using quantum states $|L,L_z\rangle=|L,L\rangle$ with $L_z=L$.
According to Wigner-Eckart
\begin{equation}
\label{we}
\langle L|| Q^{(2)} || L \rangle = \left. \langle L, L| Q^{(2)}_0 | L, L \rangle \right/ \left(\begin{array}{ccc}
L & 2 & L\\
-L & 0 & L
\end{array}\right),
\end{equation}
where $Q^{(2)}_0 = Q_{zz} = 2 (p_z^2 - p^2/3)$, and the 3j-symbol is
\begin{eqnarray}
\left(\begin{array}{ccc}
L & 2 & L\\
-L & 0 & L
\end{array}\right) = \sqrt{\frac{L(2 L - 1)}{(L+1)(2 L+1)(2 L+ 3)}}.
\nonumber
\end{eqnarray}
To calculate $\langle L, L| Q^{(2)}_0 | L, L \rangle$
we use the explicit form of $p_z^2$  in spherical coordinates:
\begin{eqnarray}
p_z^2 = - \left( \cos^2\theta \cdot \partial_{rr} + \frac{1}{r} \partial_r \cdot (\sin^2 \theta - \sin(2 \theta) \cdot \partial_\theta) + \right. \nonumber \\ \left. + \frac{1}{r^2} (\sin^2\theta \cdot \partial_{\theta \theta} + \sin(2 \theta) \cdot \partial_\theta)\right)\ . \nonumber
\end{eqnarray}
The spherical harmonic $Y_{LL} = |L, L \rangle$ has the following well-known form
\begin{eqnarray}
Y_{LL}(\theta, \phi) = (-1)^L e^{i L \phi} \sqrt{\frac{(2L+1)!!}{4 \pi (2 L)!!}} \sin^L\theta \ . \nonumber
\end{eqnarray}
Performing the straightforward integration over the angles 
we find the matrix element
\begin{eqnarray}
\langle L, L| Q^{(2)}_0 | L, L \rangle = \frac{4 L}{3 (2 L+3)} \left( \partial_{rr} + \frac{2}{r} \partial_r - \frac{L(L+1)}{r^2} \right) \nonumber
\end{eqnarray}
and hence using (\ref{we}) we find the reduced matrix element.
The procedure to find the offdiagonal matrix element 
$\langle L + 2|| Q^{(2)} || L \rangle$ is absolutely similar.
All in all this gives
	\begin{eqnarray}
\label{eqd}
&&	\langle L|| Q^{(2)} || L \rangle = 
\frac{4}{3} \sqrt{\frac{L(L+1)(2L+1)}{(2L - 1)(2L + 3)}}\\
&&\hspace{60pt} \times\left(\partial_{rr} + \frac{2}{r} \partial_r - \frac{L(L+1)}{r^2} \right)\nonumber \\
&&	\langle L + 2|| Q^{(2)} || L \rangle = 
-2  \sqrt{\frac{2}{3}\frac{(L+1)(L+2)}{2L + 3}} \nonumber\\
&&\hspace{70pt}\times\left( \partial_{rr} - \frac{2 L + 1}{r} \partial_r + \frac{L(L+2)}{r^2} \right) \nonumber\\
&&	\langle L - 2|| Q^{(2)} || L \rangle = 
-2  \sqrt{\frac{2}{3}\frac{L(L-1)}{2L - 1}} \nonumber\\
&&\hspace{70pt}\times\left( \partial_{rr} + \frac{2 L + 1}{r} \partial_r + \frac{L^2 - 1}{r^2} \right) \ . \nonumber
	\end{eqnarray}
The reduced matrix element of $\tau^{(2)}$ is presented in Eq.(\ref{rm}).
Substituting the wave function~(\ref{wavef}) in the 
Schr$\ddot{\mathrm o}$dinger equation with Hamiltonian (\ref{LutSP1}) 
and taking the projections on spherical harmonics $|L, S, J, M \rangle$ 
we finally obtain equations (\ref{jl}), (\ref{radsyst11}), and 
(\ref{radsyst12}) for radial functions. 
When calculating matrix elements of $Q_{ij} \tau_{ij}$ in Eq.(\ref{LutSP1})
one has to remember the following relation between the Cartesian dot product
and the spherical dot product (\ref{qtau}),
$Q_{ij} \tau_{ij} = \frac{3}{2} (Q \cdot \tau)$.


\begin{thebibliography}{10}
	\bibitem{luttinger} J. M. Luttinger, \emph{Phys. Rev.} {\bf 102}, 4 (1956).
	\bibitem{baldereschi_sph} A. Baldereschi, N. O. Lipari, \emph{Phys. Rev. B} {\bf 8}, 2697 (1973).
	\bibitem{baldereschi_cube} A. Baldereschi, N. O. Lipari, \emph{Phys. Rev. B} {\bf 9}, 1525 (1974).
	\bibitem{baldereshi_Si_Ge} N. O. Lipari, A. Baldereschi, \emph{Solid State Commun.} {\bf 25} (1978).
	\bibitem{fiorentini} V. Fiorentini, \emph{Phys. Rev. B} {\bf 51}, 15 (1995).
	\bibitem{drachenko} O. Drachenko, H. Schneider, M. Helm, D. Kozlov, V. Gavrilenko, J. Wosnitza, J. Leotin, \emph{Phys. Rev. B} {\bf 84}, 245207 (2011).
	\bibitem{Said} M. Said, M. A. Kanehisha, \emph{Phys. Stat. Sol. B} {\bf 157}, 311 (1990).
	\bibitem{xia} J.-B. Xia, \emph{Phys. Rev. B} {\bf 40}, 12 (1989).
	\bibitem{Schmitt} W. O. G. Schmitt, E. Bangert, G. Landwehr, \emph{J. Phys. Condens. Matter} {\bf 3} (1991).
	\bibitem{malyshev} A. V. Malyshev, \emph{Phys. Solid State} {\bf 42}, 1 (2000).
	
	\bibitem{atzmuller} R. Atzm$\ddot {\mathrm u}$ller, M. Dahl, J. Kraus, G. Schaack, J. Schubert, \emph{J. Phys. Condens. Matter} {\bf 3} (1991).
	\bibitem{linnarsson} M. Linnarsson, E. Janz$\acute {\mathrm e}$n, B. Monemar, \emph{Phys. Rev. B} {\bf 55}, 11 (1997).
	\bibitem{lewis_gaas} R. A. Lewis, Y.-J. Wang, M. Henini,  \emph{Phys. Rev. B} {\bf 67}, 235204 (2003).
	
	\bibitem{fisher} P. Fisher, G. J. Takacs, R. E. M. Vickers, A. D. Warner, \emph{Phys. Rev. B} {\bf 47}, 19 (1993).
	\bibitem{baker} R. J. Baker, P. Fisher, C. A. Freeth, D. S. Ryan, R. E. M. Vickers, \emph{Solid State Commun.} {\bf 93}, 5 (1995).
	\bibitem{vickers} P. Fisher, R. E. M. Vickers, \emph{Solid State Commun.} {\bf 100}, 4 (1996).
	\bibitem{prakabar} P. C. J. Prabakar, R. E. M. Vickers, P. Fisher, \emph{Solid State Commun.} {\bf 107}, 2 (1998).
	\bibitem{wang} R. E. M. Vickers, R. A. Lewis, P. Fisher, Y.-J. Wang, \emph{Phys. Rev. B} {\bf 77}, 115212 (2008).
	
	\bibitem{ramdas} A. K. Ramdas, S. Rodriguez, \emph{Rep. Prog. Phys.} {\bf 44} (1981).
	\bibitem{kopf} A. K$\ddot {\mathrm o}$pf, K. Lassmann, \emph{Phys. Rev. Lett.} {\bf 69}, 10 (1992).
	\bibitem{heijden} J. van der Heijden, J. Salfi, J. A. Mol, J. Verduijn, G. C. Tettamanzi, A. R. Hamilton, N. Collaert, S. Rogge, \emph{Nano Lett.} {\bf 14} (2014).
\bibitem{Winkler}
R. Winkler, \emph{Spin-Orbit Coupling Effects in Two-Dimensional Electron and Hole Systems} (Springer-Verlag Berlin Heidelberg, 2003).
\bibitem{Landau}L. D. Landau and E. M. Lifshitz, \emph{Quantum Mechanics Non-relativistic Theory} (Pergamon Press, 1965).
\bibitem{Uli}
Y. Komijani, M. Csontos, I. Shorubalko, U. Z$\ddot {\mathrm u}$licke, T. Ihn, K. Ensslin, D. Reuter, and A. D. Wieck,
\emph{Europhys. Lett.} {\bf 102}, 37002 (2013).  
\end{thebibliography}
\end{document}